\documentclass[aps,pre,twocolumn,groupedaddress]{revtex4}
\usepackage{graphicx}
\usepackage{amsmath}

\begin{document}

\title{Duality with real-space renormalization and its application to bond percolation}
\author{Masayuki Ohzeki}
\date{\today}

\address{Department of Systems Science,
Graduate School of Informatics, Kyoto University, Yoshida-Honmachi,
Sakyo-ku, Kyoto, 606-8501, Japan}

\begin{abstract}
We obtain the exact solution of the bond-percolation thresholds with inhomogenous probabilities on the square lattice.
Our method is based on the duality analysis with real-space renormalization, which is a profound technique invented in the spin-glass theory.
Our formulation is a more straightforward way compared to the very recent study on the same problem [R. M. Ziff, {\it et. al.}, J. Phys. A: Math. Theor. 45 (2012) 494005]. 
The resultant generic formulas from our derivation can give several estimations for the bond-percolation thresholds on other lattices rather than the square lattice.
\end{abstract}
\maketitle
\section{Introduction}

Forest fire happens suddenly and spreads out rapidly.
In order to save the forest itself, living animals, humans and their community there, it is important to resolve a naive question: how can we prevent the fire spread through the whole system?
In the present study we take an associated mathematical problem, namely {\it percolation} \cite{Stauffer1994}.
The percolation is a very simple but ubiquitous problem, which is closely related to the phenomenon involved in the formation of long-range connectivity in systems, as well as forest fire as exemplified above.
For instance, it provides rich comprehensions for numerous practical issues including conductivity in composite materials, infectious disease, flow through porous media, and polymerization.
In the present study we restrict ourselves to the case of the bond percolation problem, where each bond to connect both ends on the system is selected in a stochastic manner.
The bond percolation is a typical instance of the cooperative phenomena, with which highly skillful techniques are essential to deal.
Nevertheless very simple formulas have been expected to hold for the bond-percolation thresholds at which giant clusters over the whole system appear.
The key is a particular symmetry embedded in the system, namely the duality.

In classical spin models, the duality is known to be a hidden symmetry between the partition functions in low and high temperatures.
This symmetry allows us to identify the locations of the critical points for various spin models such as the Ising and Potts models \cite{Kramers1941,Wu1976}.
In the present study we employ the duality in order to assess the bond-percolation threshold, since $q$-state Potts model can be mapped to the bond-percolation problem in the limit of $q\to 1$ \cite{Wu1982,Nishimori2011}.
The special symmetry of the square lattice, namely self-duality, yields the exact solution of the bond-percolation threshold in the case with a homogenous probability on each bond.
Even for the case without self-duality, we can perform the duality analysis to obtain the bond-percolation thresholds in several cases in conjunction with another technique, namely the star-triangle transformation \cite{Wu1982}.
 
In the present study, we generalize the star-triangle transformation to the case on the square lattice.
We apply the generalized technique, namely the duality analysis with real-space renormalization, to the inhomogenous case on the square lattice.
The resultant equation to provide the bond-percolation thresholds coincides with that proposed by Wu \cite{Wu1979}.
Very recent work performed by Ziff, {\it et. al.} has proved its validity by combination of the several profound results \cite{Ziff2012}.
Our technique provides a more straightforward way to derive the exact formula on the critical manifolds of the bond-percolation thresholds without any other ingredients to support our analysis.
Moreover, we give explicit forms of several generic formulas depending on the structure of the unit cell forming the lattice.
The basis of our technique comes from the different stream of study on random spin systems, in particular spin glasses.
The straightforward rederivation of the existing equalities in the different context implies existence of close connection between different realms, bond-percolation problems and spin glasses.

The paper is organized as follows.
In the next section, we review the conventional duality and the star-triangle transformation for convenience.
The third section demonstrates the duality with real-space renormalization to the inhomogenous case on the square lattice.
In \S 4 we find the resultant generic formulas for the bond-percolation thresholds and compare our results to the very recent studies.
In the last section, we conclude our study.
\section{Conventional analysis}

\begin{figure}[tbp]
\begin{center}
\includegraphics[width=70mm]{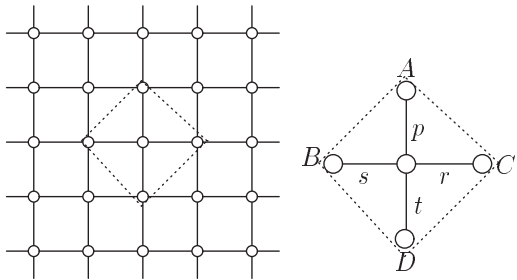}
\end{center}
\caption{The unit cell of the inhomogenous bond-percolation problem on the square lattice.
The unit cell is a part of the square lattice, which is covered with the dashed lines. 
The assigned values $p$, $r$, $s$ and $t$ are the probabilities to connect both ends on each bond on the unit cell.}
\label{fig1}
\end{figure}

Our analysis is based on the duality \cite{Kramers1941,Wu1976}, which is the simplest way to estimate the bond-percolation thresholds.
We consider the bond-percolation thresholds for the lattice consisting of repetition of the unit cell as in Fig. \ref{fig1}.
Let us define $p$, $r$, $s$, and $t$ as the inhomogenous probabilities to connect both ends of the assigned bonds.
The conventional duality analysis can lead to the bond-percolation thresholds for the homogenous case $p=r=s=t$ on the square lattice.
In addition to the duality, the star-triangle transformation gives the bond-percolation thresholds for the inhomogenous case on the triangular and hexagonal lattices.
First, let us review the conventional duality for convenience.

\subsection{Duality}

We consider the $q$-state Potts model with the following Hamiltonian,
\begin{equation}
H = -  \sum_{\langle ij \rangle} J_{ij} \delta\left(\phi_i - \phi_j\right),
\end{equation}
where $J_{ij}$ is the strength of interactions and takes different values as $J_p$, $J_r$, $J_s$, and $J_t$, which will correspond to the probability assigned on the bonds.
The summation is taken over all bonds, $\delta(x)$ is Kronecker's delta, and $\phi_i$ stands for the spin direction taking $0,1,\cdots$, and $q-1$.
Let us estimate the critical point of the $q$-state Potts model, since it corresponds to the bond-percolation threshold in the limit of $q \to 1$ \cite{Wu1982,Nishimori2011}. 

We here assume the homogeneous case $J=J_p=J_r=J_s=J_t$.
The duality exploits an inherent symmetry embedded in the partition function with the inverse temperature $\beta$ as $Z=\sum_{\phi_i}\prod_{\langle ij \rangle}\exp(\beta J \delta(\phi_i - \phi_j))=\sum_{\phi_i}\prod_{\langle ij \rangle}(1+ v\delta(\phi_i - \phi_j))$, where $v=\exp(\beta J)-1$ \cite{Kramers1941}.
Two different approaches to evaluate the partition function, the low- and high-temperature expansions, can be related to each other by the $q$-component discrete Fourier transformation for the local part of the Boltzmann factor, namely edge Boltzmann factor $x_k = 1+ v\delta(k)$ \cite{Wu1976}.
Specifically, each term in the low-temperature expansion can be expressed by $x_k$, while the high-temperature one is written in the dual edge Boltzmann factor $x_l^*=\sum_{k} x_k \exp(i2\pi k l/q)/\sqrt{q}$.
As a result, we obtain a double expression of the partition function by use of two different edge Boltzmann factors as
\begin{equation}
Z(x_0,x_1,\cdots) = q^{N_S-\frac{N_B}{2}-1}Z^*(x^*_0,x^*_1,\cdots) \label{duality0}.
\end{equation}
where $Z^*$ is the partition function on a dual lattice.
Here $N_S$ and $N_B$ denote the numbers of sites and plaquettes, respectively.
The unity in the power of $q$ can be ignored in the following analysis.
We obtain another system on the dual graph, on which each site on the original lattice exchanges with each plaquette on the dual one and vice versa, after the dual transformation through the $q$-component discrete Fourier transformation.
When the dual lattice is the same as the original one, the system holds self-duality.
For instance, the square lattice is the case.
Then we can regard $Z^*(x^*_0,x^*_1,\cdots)$ as $Z(x^*_0,x^*_1,\cdots)$ and can obtain the exact value of the critical point by the duality.
We restrict ourselves to the case on the square lattice.
Notice that $N_B/2=N_S$ on the square lattice.
Let us extract the principal Boltzmann factors with edge spins parallel $x_0$ and $x_0^*$ from both sides of Eq. (\ref{duality0}) as
\begin{equation}
(x_0)^{N_B}z(u_1,u_2,\cdots) = (x^*_0)^{N_B}z(u^*_1,u^*_2,\cdots),
\end{equation}
where $z$ is the normalized partition function $z(u_1,u_2,\cdots)=Z/(x_0)^{N_B}$ and $z(u^*_1,u^*_2,\cdots)=Z/(x^*_0)^{N_B}$.
We here define the relative Boltzmann factors $u_k = x_k/x_0=1/(1+v)$ and $u_k^*= x^*_k/x^*_0=v/(q+v)$.
The well-known duality relation can be obtained by rewriting $u^*_k$ in the same form as $u_k$ by use of $v^*$ as $v/(q+v) = 1/(1+v^*)$, namely $v^*=q/v$.
Notice that the quantity $v^*$ has a different parameter $K^*$ from the original coupling $K=\beta J$, which implies transformation of the temperature.
We obtain the exact value of the critical temperature from the fixed point condition $v^2_c=q$ under the assumption that a unique transition undergoes in the system.
The limit $q \to 1$ can then give the bond-percolation threshold in the homogenous case $p_c=p=r=s=t$ on the square lattice through $p_c = v_c/(1+v_c)$, namely $p_c=1/2$ \cite{Wu1982,Nishimori2011}.
We can also derive the critical point by the following simple equality
\begin{equation}
x_0 = x_0^*.\label{MCP_duality}
\end{equation}
Indeed this equality gives $v_c=1$, namely $p_c=1/2$.

\begin{figure}[tbp]
\begin{center}
\includegraphics[width=70mm]{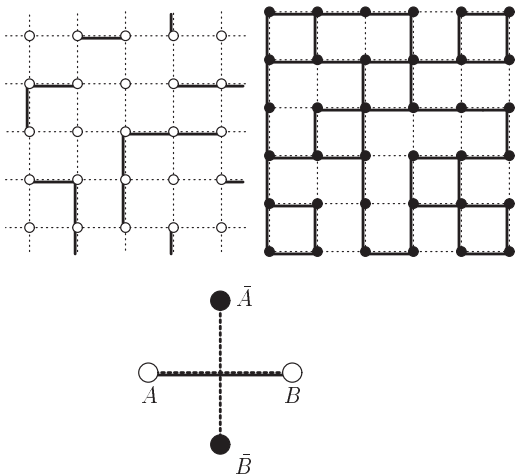}
\end{center}
\caption{Duality relation of the bond-percolation problem.
The dotted line denotes the disconnected bond.
The bold line represents the connected bond.
The white circles denote the original sites.
The black circles represent the dual sites (original plaquettes).
}
\label{fig2}
\end{figure}

For the case without self-duality, we can find an important relation from $v^*=q/v$.
We can relate the probability assigned on the bond on the original lattice to that on the dual one as $p^* = 1-p$ in the limit $q \to 1$ \cite{Wu1982,Kesten1982}.
In other words, the probability $p$ that both ends are connected on the original lattice is transformed into the disconnected probability on the dual lattice as $1-p^*$ and vise versa.
We can rewrite this fact in terms of the relationship of the connectivity as
\begin{equation}
P(AB) = P(\bar{A}|\bar{B}),\label{PPdual}
\end{equation}
where the quantity on the left-hand side expresses the probability that $A$ and $B$ are connected, and that on the right-hand side stands for the probability that $\bar{A}$ and $\bar{B}$ are disconnected.
The end points $A$ and $B$ in Fig. \ref{fig2} denote the sites on the original lattice.
On the other hand, $\bar{A}$ and $\bar{B}$ represent the sites on the dual lattice.
Then the bond percolation threshold for the homogenous case on the square lattice can be represented by the following equality
\begin{equation}
P(AB) = P(A|B).\label{PP0}
\end{equation}

\subsection{Star-triangle transformation}

\begin{figure}[tbp]
\begin{center}
\includegraphics[width=70mm]{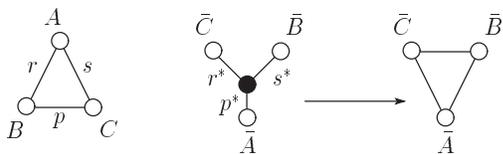}
\end{center}
\caption{The inhomogeneous bond-percolation problem on the triangular and hexagonal lattices.
The assigned values $p$, $r$ and $s$ are the connected probabilities assigned on each bond on the unit cell.
On the hexagonal lattice, we put the dual probabilities $p^*$, $r^*$ and $s^*$, which are obtained after the dual transformation.
The black circle on the hexagonal lattice represents the internal site we sum over in the star-triangle transformation.
}
\label{fig3}
\end{figure}

Let us consider the case on the triangular lattice.
We here remove the homogeneous restriction that we impose above.
We deal with the bond-percolation problem with the inhomogenous probabilities on the triangular lattice as depicted in Fig. \ref{fig3}.

The dual transformation changes the triangular lattice into the hexagonal lattice.
Then we cannot perform the same analysis as that in the case on the square lattice.
We employ another technique to relate the hexagonal lattice to the original triangular lattice.
This can be achieved by the partial summation over internal spins at the down-pointing (up-pointing) star on the hexagonal lattice, namely star-triangle transformation \cite{Wu1982}.
Then we can transform the partition function on the hexagonal lattice into that on another triangular lattice, namely $Z^*(x^*_0,x^*_1,\cdots)=Z(x^{*({\rm tr})}_0,x^{*({\rm tr})}_1,\cdots)$ in Eq. (\ref{duality0}).
We here use the renormalized-edge Boltzmann factor $x^{*({\rm tr})}_k$ defined as
\begin{equation}
x_k^{*({\rm tr})} = \frac{1}{\sqrt{q}} \sum_{\phi_0} \prod_{i}\left\{\frac{v_i}{\sqrt{q}}\left(1 + \delta(\phi_i- \phi_0)\frac{q}{v_i}\right)\right\},\label{RBS}
\end{equation}
where the product runs over $i=p,r$, and $s$ for the three bonds on the unit cell of the hexagonal lattice, namely the down-pointing (up-pointing) star.
We here assume the inhomogeneous system with $v_i = \exp(\beta J_i)-1$.
We take the summation over the internal spin $\phi_0$ denoted by the black circle on the unit cell as in Fig. \ref{fig3}.
The coefficient $1/\sqrt{q}$ comes from that in front of the partition function on the right-hand side of Eq. (\ref{duality0}).
Notice that $N_S$ is the same as the number of down-pointing (up-pointing) triangles $N_{\rm tr}$ on the triangular lattice, and $N_B=3N_S$.
The subscript $k$ denotes the configuration of the edge spins $\{\phi_{l=p,r,s}\}$ on the unit cell.
On the other hand, we rewrite the original partition function in terms of the product of the edge Boltzmann factors as
\begin{equation}
x_k^{({\rm tr})} = \prod_{i}\left(1 + \delta(\phi_i- \phi_0)v_i\right).\label{RBT}
\end{equation}
The double expression of the partition function can be written as 
\begin{equation}
Z(x^{({\rm tr})}_0,x^{({\rm tr})}_1,\cdots) = Z^*(x^{*({\rm tr})}_0,x^{*({\rm tr})}_1,\cdots) \label{duality_th}.
\end{equation}
Similarly, let us extract the renormalized-principal Boltzmann factors with edge spins parallel $x^{({\rm tr})}_0$ and $x_0^{*({\rm tr})}$ from both sides of Eq. (\ref{duality_th}) as
\begin{eqnarray}\nonumber
&& \{x^{({\rm tr})}_0\}^{N_{\rm tr}}z^{({\rm tr})}(u^{({\rm tr})}_1,u^{({\rm tr})}_2,\cdots) \\
&&= \{x_0^{*({\rm tr})}\}^{N_{\rm tr}}z^{({\rm tr})}(u^{*({\rm tr})}_1,u^{*({\rm tr})}_2,\cdots).
\end{eqnarray}
Notice that the number of the down-pointing (up-pointing) stars is the same as $N_{\rm tr}$.
and $z^{({\rm tr})}$ is the normalized partition function $z^{({\rm tr})}(u^{({\rm tr})}_1,u^{({\rm tr})}_2,\cdots)=Z/(x^{({\rm tr})}_0)^{N_{\rm tr}}$ and $z^{({\rm tr})}(u^{*({\rm tr})}_1,u^{*({\rm tr})}_2,\cdots)=Z/(x^{*({\rm tr})}_0)^{N_{\rm tr}}$.
We here define the renormalized-relative Boltzmann factors $u^{({\rm tr})}_k = x^{({\rm tr})}_k/x^{({\rm tr})}_0$ and $u_k^{*({\rm tr})}= x^{*({\rm tr})}_k/x^{*({\rm tr})}_0$.
Similarly to the case on the square lattice, we put the simple equality as
\begin{equation}
x^{({\rm tr})}_0 = x^{*({\rm tr})}_0.\label{MCP_th}
\end{equation}
This equality yields the critical manifold of the $q$-state Potts model as detailed in Appendix \ref{AP0}.
By taking the limit $q \to 1$, we obtain the equality for the bond-percolation thresholds on the triangular lattice as
\begin{equation}
T(p,r,s) = 0,\label{Teq0}
\end{equation}
where
\begin{equation}
T(p,r,s) = prs - p -r -s +1. \label{Teq}
\end{equation}
If we perform the dual transformation on this equality, we find the solution for the bond-percolation thresholds on the hexagonal lattice as
\begin{equation}
H(p^*,r^*,s^*) = 0, \label{Heq0}
\end{equation}
where
\begin{equation}
H(p,r,s) = prs - rp -rs-ps +1.\label{Heq}
\end{equation}
As shown above we can obtain the exact solution of the bond-percolation thresholds for several cases through the duality and the technique in conjunction with the star-triangle transformation.

\section{Duality with real space renormalization}

The duality with the star-triangle transformation, which is the partial summation of the unit cell on the hexagonal lattice, leads to the exact solution for the bond-percolation thresholds on the triangular and hexagonal lattices as in (\ref{Teq0}) and (\ref{Heq0}).
Let us develop the similar analysis to the successful case on the triangular and hexagonal lattices.

\begin{figure}[tbp]
\begin{center}
\includegraphics[width=90mm]{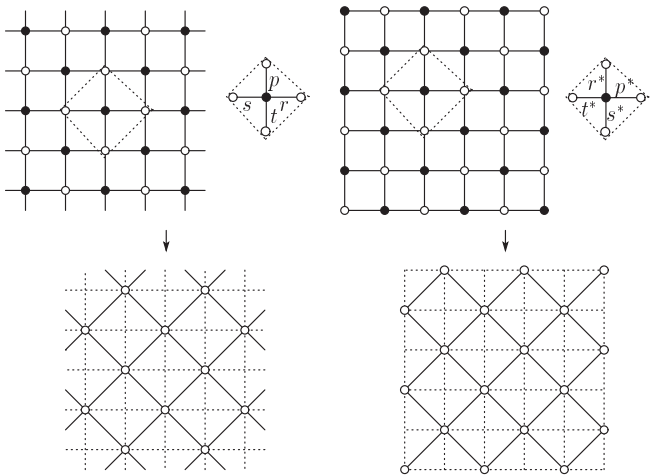}
\end{center}
\caption{The unit cell with four bonds for the inhomogenous case on the square lattice.
The black circle denotes the internal spins that we sum over, while the white ones are fixed as $\phi_i = 0$.}
\label{fig4}
\end{figure}

We start from Eq. (\ref{duality0}) for the case on the square lattice.
Notice that the edge Boltzmann factor is not enough to express the local property of the inhomogeneous system.
Thus we consider to use the renormalized-edge Boltzmann factor inspired by the star-triangle transformation.
We take the square unit cell consisting of four bonds from both of the original and dual square lattices as in Fig. \ref{fig4}.
Let us take the product of the edge Boltzmann factors and perform the summation over the internal spin.
The resultant quantity is written as
\begin{equation}
x_k^{({\rm sq})} = \sum_{\phi_0} \prod_{i=p,r,s,t}\left(1+v_i\delta(\phi_i- \phi_0)\right).\label{RB1}
\end{equation}
Similarly, we obtain the dual renormalized-edge Boltzmann factor as
\begin{equation}
x_k^{*({\rm sq})} = \sum_{\phi_0} \prod_{i=p,r,s,t}\left\{\frac{v_i}{\sqrt{q}}\left(1 + \delta(\phi_i- \phi_0)\frac{q}{v_i}\right)\right\}.\label{RB2}
\end{equation}
We can then rewrite the relation obtained by the conventional duality (\ref{duality0}) as
\begin{equation}
Z(x^{({\rm sq})}_0,x^{({\rm sq})}_1,\cdots) = Z(x^{*({\rm sq})}_0,x^{*({\rm sq})}_1,\cdots) \label{duality1}.
\end{equation}
We extract the renormalized-principal Boltzmann factors $x^{({\rm sq})}_0$ and $x_0^{*({\rm sq})}$ with all edge spins on the unit cell parallel as
\begin{eqnarray}\nonumber
&&(x^{({\rm sq})}_0)^{N_B/4}z^{({\rm sq})}(u^{({\rm sq})}_1,u^{({\rm sq})}_2,\cdots) \\
&& \quad = (x^{*({\rm sq})}_0)^{N_B/4}z^{({\rm sq})}(u^{*({\rm sq})}_1,u^{*({\rm sq})}_2,\cdots),
\end{eqnarray}
where $z^{({\rm sq})}$ is the normalized partition function but $z^{({\rm sq})}(u^{({\rm sq})}_1,u^{({\rm sq})}_2,\cdots)=Z/(x^{({\rm sq})}_0)^{N_B/4}$ and $z(u^{*({\rm sq})}_1,u^{*({\rm sq})}_2,\cdots)=Z/(x^{*({\rm sq})}_0)^{N_B/4}$.
We here define the renormalized-relative Boltzmann factors $u^{({\rm sq})}_k = x^{({\rm sq}))}_k/x^{({\rm sq})}_0$ and $u_k^{*({\rm sq})}= x^{*({\rm sq})}_k/x^{*({\rm sq})}_0$.
Then we impose the following equation to identify the location of the critical point
\begin{equation}
x^{({\rm sq})}_0 = x^{*({\rm sq})}_0.\label{MCP1}
\end{equation}
The direct evaluation of this equality in the leading order of $\epsilon$ where $q = 1 + \epsilon$ gives the formula for the bond-percolation thresholds, as detailed in Appendix \ref{AP1},
\begin{equation}
\prod_{i} (1+v_i) C(p,r,s,t) = 0.\label{Per1}
\end{equation}
where
\begin{eqnarray}\nonumber
& & C(p,r,s,t) = 1 -pr -ps -rs -pt -rt -st \\
& & \quad +prs +prt+rst+pst.\label{Ceq}
\end{eqnarray}
It is reasonable that a unique transition undergoes if we tune the temperature for the inhomogeneous interactions as $J_p$, $J_r$, $J_s$, and $J_t$, which correspond to the probabilities of the bond-percolation problem in the limit $q \to 1$.
Therefore the singularity of the free energy should be unique for change of the temperature.
The duality can then identify the location of the critical point.
Therefore we conclude that $C(p,r,s,t) = 0$ gives the exact bond-percolation thresholds for the inhomogeneous case on the square lattice.

The equality $C(p,r,s,t) = 0$ was originally conjectured \cite{Wu1979}, confirmed numerically with high precision and derived in a different way \cite{Scullard2008}.
The proof of validity of Eq. (\ref{Ceq}) has been very recently established \cite{Ziff2012}.
It is not simple to show the validity of Eq. (\ref{Ceq}), since it is based on the indirect analysis via considerations of the bond-percolation problem on different lattices.
The present analysis demonstrates the more straightforward analysis.
Without recourse to the duality, the real-space renormalization group analysis can give the exact bond-percolation thresholds for the homogenous case but fails into the approximations for the inhomogenous case \cite{Raynolds1978,Nakanishi1981}.
By virtue of the duality, we can here find the exact answer for the critical point while we stand on the fixed point of the renormalization group.

\begin{figure}[tbp]
\begin{center}
\includegraphics[width=50mm]{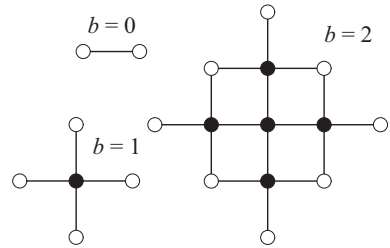}
\end{center}
\caption{The cluster for the duality with real-space renormalization on the square lattice.
The white circles denote the fixed spins for evaluation of the principal Boltzmann factors.
The black circles express the spins we trace over (similarly to the star-triangle transformation) to obtain the principal Bolztmann factors.}
\label{fig5}
\end{figure}

The duality with real-space renormalization as shown above is essentially the same as the profound technique in the analysis of the random spin system, in particular spin glasses \cite{Ohzeki2008,Ohzeki2009a}.
For several models in the random spin system, the dual transformation cannot relate the original system to the same one with a different temperature as the case for the $q$-state Potts model despite existence of self-duality of the lattice.
In these cases, we recover the self-duality of the random spin models via real-space renormalization over larger range beyond the unit cell, namely summation over several internal spins by taking larger size of the cluster in order to find the correct fixed point in a relatively wide space of parameters as well as the temperature.
Then we impose the following condition to estimate the critical point similarly to the conventional analysis by the duality (\ref{MCP_duality}), its combination with the star-triangle transformation (\ref{MCP_th}), and the above analysis (\ref{MCP1}) \cite{Ohzeki2008,Ohzeki2009a}
\begin{equation}
x_0^{(b)}=x_0^{*(b)}, \label{MCP_cluster}
\end{equation}
where $x_0^{(b)}$ and $x_0^{*(b)}$ are renormalized principal and dual Boltzmann factors on the cluster.
The size of the cluster is denoted by $b$ ($b=0$ means the simple duality without renormalization).
It is examined that, if $b$ is taken to be a large value, we can recover the self-duality following the concept of renormalization \cite{Raynolds1977,Raynolds1980}, and Eq. (\ref{MCP_cluster}) can give a precise estimation of the critical point \cite{Ohzeki2008}.
When we deal with the random spin system on the square lattice, the estimation by setting $b=1$ (four-bond cluster as depicted in Fig. \ref{fig5}) often attains satisfiable precision.
If one wishes to enhance the precision,  the systematic improvement is achievable by increase of $b$ (i. e. $b=2$ 16-bond cluster as in Fig, \ref{fig5}). 
Indeed we can also estimate the bond-percolation thresholds from the critical points via analysis on the bond-dilution Ising model, which is a typical model in the random spin system.
In Appendix \ref{AP2}, we demonstrate the rederivation of the formula $C(p,r,s,t) = 0$ through the duality analysis with real-space renormalization for the bond-dilution Ising model with inhomogenous distribution.
Equation (\ref{MCP_cluster}) yields the formula $C(p,r,s,t) = 0$ for the bond-percolation threshold for the inhomogeneous case on the square lattice.
Below we examine the obtained result through the duality with real-space renormalization in several points of view.
Beyond the case on the square lattice, we try to apply several similar systems.

\section{Generic formulas on bond percolation thresholds}

When we analyze the homogenous case on the square lattice, the formula of the bond-percolation threshold is a relation on the unit cell with the two-terminal structure (a single bond) as in Eq. (\ref{PP0}).
On the other hand, the unit cell where we perform the analysis consists of the three-terminal structure for the case on the triangular and hexagonal lattice as depicted in Fig. \ref{fig3}.
As in this case, for the lattice consisting of repetition of the three-terminal unit cell, the generic formula of the bond-percolation thresholds is known to be \cite{Ziff2006}.
\begin{equation}
P(ABC) = P(A|B|C), \label{PP1}
\end{equation}
where the quantity on the left-hand side expresses the probability that the end points $A$, $B$, and $C$ on the three-terminal unit cell are all connected, and that on the right-hand side stands for the probability that none of $A$, $B$, and $C$ are connected.

We obtain the critical manifold (\ref{Teq0}) for the bond-percolation thresholds on the triangular lattice from the above formula (\ref{PP1}).
We here demonstrate the reduction to Eq. (\ref{Teq0}) from Eq. (\ref{PP1}).
In the case on the triangular lattice, let us write down all terms included in $P(ABC)$ as
\begin{eqnarray}\nonumber
P(ABC) &=& prs + pr(1-s) + p(1-r)s + (1-p)rs \\
&=& pr+sp+rs - 2 prs.
\end{eqnarray}
On the other hand, the probability that none of the end points are connected is
\begin{eqnarray}\nonumber
P(A|B|C) &=& (1-p)(1-r)(1-s)\\ \nonumber
&=& 1-p-r-s+pr+rs+sp-prs.\\
\end{eqnarray}
Thus we can obtain Eq. (\ref{Teq0}) from Eq. (\ref{PP1}).

In addition, on the hexagonal lattice, Eq. (\ref{PP1}) can be reduced to Eq. (\ref{Heq0}).
On the hexagonal lattice, the left-hand side of Eq. (\ref{PP1}) is written as
\begin{eqnarray}\nonumber
P(ABC) &=& prs.
\end{eqnarray}
The right-hand side can be given as
\begin{eqnarray}\nonumber
P(A|B|C) &=& (1-p)(1-r)(1-s)+p(1-r)(1-s)\\ \nonumber
&&\quad +(1-p)r(1-s)+(1-p)(1-r)s \\ 
&=& 1-pr-rs-sp+2prs.
\end{eqnarray}
Equation (\ref{PP1}) reproduces Eq. (\ref{Heq0}).

We here give a generic formula for several lattices consisting of repetition of the four-terminal unit cell as in Fig. \ref{fig6} (left).
The analysis as detailed in Appendix \ref{AP2} provides the generic formula for the four-terminal unit cell as
\begin{eqnarray}\nonumber
&&P(ABCD) + P(BCD|A) +P(ACD|B) \\ \nonumber
&&\quad +P(ABD|C)+P(ABC|D) = 
P(A|B|C|D), \\ \label{PP2}
\end{eqnarray}
where $P(BCD|A)$ is the probability that $BCD$ connects with each other while $A$ is disconnected, and the other quantities follow the same manner.
We can reproduce Eq. (\ref{PP1}) by reduction to the three-terminal unit cell (in particular hexagonal lattice) by removing the single bond from four bonds as in Fig. \ref{fig6} (left).
It means that all connected probabilities to $D$ vanish as $P(ABCD)=0$ since $D$ can not be connected, and we omit dependence on $D$ for the disconnected probability with $D$ as $P(ABC|D)=P(ABC)$.
\begin{figure}[tbp]
\begin{center}
\includegraphics[width=90mm]{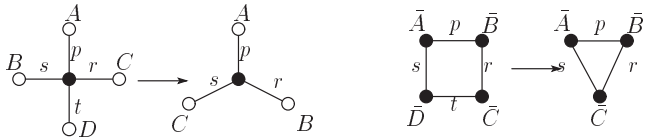}
\end{center}
\caption{Transformations into the hexagonal and triangular lattices from the square lattice.}
\label{fig6}
\end{figure}

\begin{figure}[tbp]
\begin{center}
\includegraphics[width=80mm]{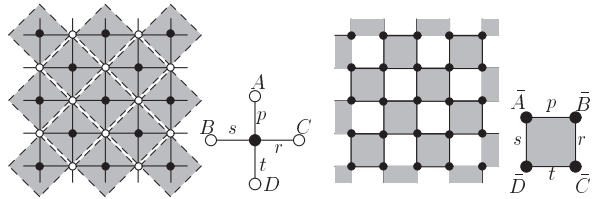}
\end{center}
\caption{Covering by the two four-terminal unit cells.
The shaded squares express the unit cells.
The left panel denotes the original square lattice, and the right one depicts the dual lattice.}
\label{fig7}
\end{figure}

For the case of the four-terminal unit cell, we can obtain another generic formula.
By the conventional duality, we relate the bond-percolation problem on the original square lattice to that on the dual square lattice through the duality relation.
The probabilities expressing connectivity of the edge sites are then changed as $P(ABCD) = P(\bar{A}|\bar{B}|\bar{C}|\bar{D})$ and $P(D|ABC)=P(\bar{D}\bar{C}|\bar{A}|\bar{B})$ similarly to Eq. (\ref{PPdual}).
Another generic formula for the four-terminal unit cell as in Fig. (\ref{fig6}) (right) can be expressed as
\begin{eqnarray}\nonumber
&&P(\bar{A}|\bar{B}|\bar{C}|\bar{D}) + P(\bar{A}\bar{B}|\bar{C}|\bar{D})+P(\bar{B}\bar{C}|\bar{D}|\bar{A}) \\ \nonumber
&&\quad +P(\bar{C}\bar{D}|\bar{A}|\bar{B})+P(\bar{D}\bar{A}|\bar{B}|\bar{C}) = 
P(\bar{A}\bar{B}\bar{C}\bar{D}), \\
\label{PP3}
\end{eqnarray}
which is detailed in Appendix \ref{AP2}.
Here let us again reduce the above equality to the case of the three-terminal unit cell.
This can be achieved by eliminating the terms associated with the disconnected probabilities with $\bar{D}$ since they are always connected as in Fig. \ref{fig6} (right).
In addition we omit the dependence on $D$.
Then Eq. (\ref{PP3}) recovers $P(\bar{A}\bar{B}\bar{C}) = P(\bar{A}|\bar{B}|\bar{C})$, namely Eq. (\ref{PP1}).
The difference between  the four-terminal unit cells associated with Eqs. (\ref{PP2}) and (\ref{PP3}) comes from the tiling manner to cover the whole lattice.
The former case is the full tiling of the unit cell.
On the other hand, the latter case is the checker-board tiling as in Fig. \ref{fig7}.

\begin{figure}[tbp]
\begin{center}
\includegraphics[width=50mm]{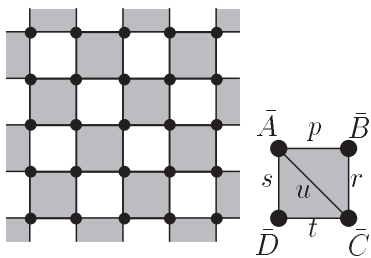}
\end{center}
\caption{The bow-tie lattice.
The shaded squares express the four-terminal unit cells.}
\label{fig8}
\end{figure}

In order to show the efficiency of the above generic formula, we take a fascinating instance of the application beyond the case of the square lattice.
By the above generic formula (\ref{PP3}), we can recover the equality for the bond-percolation thresholds on the bow-tie lattice, which is dealt with to prove the validity of Eq. (\ref{PP1}) \cite{Ziff2012}.
The four-terminal unit cell of the bow-tie lattice is shown in Fig. \ref{fig8}.
Equation (\ref{PP3}) can be then reduced to 
\begin{eqnarray}\nonumber
&&u\{prs(1-t)+pr(1-s)t+p(1-r)st+(1-p)rst\\ \nonumber
&&\quad + p(1-r)(1-s)t + p(1-r)s(1-t)\\ \nonumber 
&&\qquad + (1-p)r(1-s)t + (1-p)rs(1-t) \\ \nonumber
&&\quad \qquad + prst\} + (1-u)C(p,r,s,t) \\
&&= C(p,r,s,t) - u (1-pr-st-prst) = 0.
\end{eqnarray}
This equality has been given by combination of the results for the three-terminal unit cells by splitting of the four-terminal unit cell to two triangles as in Refs. \cite{Wierman1984,Scullard2008,Ziff2012}.
Then the combination of the duality and star-triangle transformation yields the exact solution of the bond-percolation thresholds on the bow-tie lattice.

\begin{figure}[tbp]
\begin{center}
\includegraphics[width=50mm]{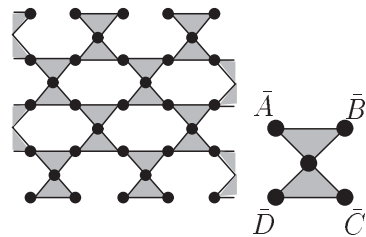}
\end{center}
\caption{The Kagom\'e lattice.
The four-terminal unit cell consists of the up-pointing and down-pointing triangles.}
\label{fig9}
\end{figure}

As another interesting but wrong instance of applications, let us apply the generic formula (\ref{PP3}) to the bond-percolation problem on the Kagom\'e lattice by considering the four-terminal unit cell with up-pointing and down-pointing triangles as in Fig. \ref{fig9}.
Here we take the homogenous case for simplicity.
We then obtain the following polynomial from the generic formula (\ref{PP3}) as
\begin{equation}
1 - 3p^2 - 6p^3 + 12 p^4 - 6p^5 + p^6 = 0.\label{KL} 
\end{equation}
The solution is $p_c = 0.524 429 71$, which is known to be an approximate estimation (Wu's conjecture \cite{Wu1976}), while a numerical evaluation gives $p_c = 0.524 405 02(5)$ \cite{Feng2008}.
The reason why the above equality yields an wrong value comes from the solvability by the duality and the stat-triangle transformation.
Our generic formula (\ref{PP3}) then fails to give the exact answer but possibly an approximate estimation when the system lacks the solvability.

However our formulation by the duality with real-space renormalization can give a more precise value of the bond-percolation threshold on the Kagom\'e lattice.
The above case is the same as that we used to find in study on random spin systems \cite{Ohzeki2008,Ohzeki2009a}.
The generic formulas as in (\ref{PP2}) and (\ref{PP3}) can be given by Eq. (\ref{MCP_cluster}) for $b=1$ through analyses of the random spin systems as detailed in Appendix \ref{AP2}.
We can obtain the precise value of the bond-percolation threshold by considering the larger cluster over the four-terminal unit cell, namely $16$-bond cluster ($b=2$) and more.
Although we here cease the discussion on the bond-percolation problem on the Kagom\'e lattice since its threshold is out of scope in the present study, we remark that several recent development on this issue.
The similar analysis is proposed  in context of the graph polynomial as demonstrated in Refs. \cite{Scullard2012a,Jacobsen2012,Scullard2012b}.
The idea is essentially based on consideration on the large cluster over the four-terminal unit cell to estimate the bond-percolation thresholds on the Kagom\'e lattice.
The method has indeed succeeded in giving very precise estimations of the bond-percolation threshold on the Kagom\'e lattice as $p_c = 0.524 405 00(1)$ \cite{Scullard2012b}.
These fact suggest that two independent methods developed in random spin systems and graph polynomial would be closely related to each other through the bond-percolation problem.
We hope that the future study reveals more clear relationship between their different realms.

\section{Conclusion}

In the present study, we rederived the exact solution for the bond-percolation thresholds in the inhomogenous case on the square lattice by use of the duality with real-space renormalization, which is a generalized analysis of the star-triangle transformation.
In addition, we obtain two different generic formulas depending on the tiling manner of the four-terminal unit cell to cover the whole lattice.
Both equalities can be reduced to the known formula for the three-terminal unit cell, which includes the triangular and hexagonal lattices.
The application of the generic formula reproduces the exact solution on the bow-tie lattice and the well-known approximate solution on the Kagom\'e lattice.
The further analysis possibly gives more precise value of the bond-percolation threshold on the Kagom\'e lattice.

The duality analysis with real-space renormalization shown in the present study is essentially the same as the special technique, which has developed in context of spin glass theory.
The method has been useful to describe the precise phase boundary \cite{Ohzeki2008,Ohzeki2009a,Ohzeki2012e}.
The straightforward rederivation of the existing results on the bond-percolation problem are given in the different context implies existence of a fascinating theoretical connection between different realms, graph polynomial and theory of spin glasses.

We emphasize high nontriviality of our results shown in the present study.
The exact solutions for finite dimensional many-body systems have been rare in spite of the long-year efforts.
However the situation begins to change by development of the duality analysis, which is found to be applicable to a relatively broad class of problems, namely spin glasses and inhomogenous percolation problems.
We hope that the duality analysis with real-space renormalization would play an essential roll to understand the nature of the many-body systems as the conventional duality proposed by Kramers and Wannier contributed to establishment of the Onsager solution \cite{Onsager1944}.

\begin{acknowledgements}
The author thanks the fruitful discussions with H. Nishimori, K. Fujii and R. M. Ziff, and is grateful to J. L. Jacobsen, T. Obuchi, and T. Hasegawa for comments on the manuscript.
This work was partially supported by MEXT in Japan, Grant-in-Aid for Young Scientists (B) No.24740263.
\end{acknowledgements}

\appendix
\section{Derivation of Eq. (\ref{Teq0}) from Eq. (\ref{MCP_th})}\label{AP0}

We evaluate Eq. (\ref{MCP_th}) in this appendix.
From the definition, we write down Eq. (\ref{MCP_th}) as
\begin{equation}
\prod_{i}(1+v_i) = \frac{1}{q^2}\left\{(q-1)\prod_{i}v_i + \prod_{i}(q+v_i)\right\}.
\end{equation}
This is the critical manifold of the $q$-state Potts model on the triangular lattice.
Let us take the leading term of $\epsilon$ of $q=1+\epsilon$ for obtaining the bond-percolation thresholds on the triangular lattice.
\begin{eqnarray}\nonumber
&&x_0^{*({\rm tr})} - x_0^{*({\rm tr})} \\ \nonumber
&&= \epsilon\prod_{i}(1+v_i)\left(-2 + \sum_i \frac{1}{1+v_i} + \prod_i \frac{v_i}{1+v_i}  \right). \\ \label{T0}
\end{eqnarray}
By rewriting each $v_i$ in terms of the probability assigned on each bond as $p = v_p/(1+v_p)$ etc., we reach 
\begin{equation}
\prod_{i}(1+v_i)T(p,r,s) = 0,
\end{equation}
which is reduced to Eq. (\ref{Teq0}).

\section{Derivation of Eq. (\ref{Ceq})}\label{AP1}

We here demonstrate the detailed evaluation of Eq. (\ref{Per1}) from Eq. (\ref{MCP1}).
We can write the difference between the left and right-hand sides of Eq. (\ref{MCP1}) by use of definition of the renormalized-edge Boltzmann factors as in Eqs. (\ref{RB1}) and (\ref{RB2}) as
\begin{eqnarray}\nonumber
&&x_0^{*({\rm sq})} -x_0^{({\rm sq})} \\ \nonumber
&&= \frac{\prod_i v_i}{q^2} \left\{q - 1 + \prod_{i}\left(1 + \frac{q}{v_i}\right)\right\}\\
&&\quad  - \left\{ q - 1 + \prod_{i}\left(1 + v_i \right)\right\}.\label{App1}
\end{eqnarray}
We take the leading term of $\epsilon$ of $q=1+\epsilon$.
In advance, we evaluate the following quantities
\begin{eqnarray}\nonumber
&&\frac{\prod_i v_i}{q^2}\left\{q - 1 + \prod_{i}\left(1 + \frac{q}{v_i}\right)\right\} \\ \nonumber
&&= \prod_i(1+v_i)\left( 1 - 2 \epsilon + \epsilon\sum_i \frac{1}{1+v_i} + \epsilon \prod_{i}\frac{v_i}{1+v_i}\right) \label{A1} \\ 
\end{eqnarray}
and
\begin{eqnarray}\nonumber
&&\left\{ q - 1 + \prod_{i}\left(1 + v_i \right)\right\} \\
&&= \prod_i(1+v_i)\left( 1 +  \epsilon \prod_{i}\frac{1}{1+v_i}\right) \label{B1}
\end{eqnarray}
Therefore Eq. (\ref{App1}) can be reduced to
\begin{eqnarray} \nonumber
&& \prod_i(1+v_i) \\ \nonumber
&& \quad \times\left\{- 2 + \sum_i \frac{1}{1+v_i} + \left(\prod_{i}v_i -1 \right)\prod_{i}\frac{1}{1+v_i} \right\} \\
&& = \prod_i(1+v_i) C(p,r,s,t).
\end{eqnarray}
We reproduce Eq. (\ref{Ceq}).

\section{Alternative way to Eq. (\ref{Ceq})}\label{AP2}

We show an alternative way to give Eq. (\ref{Ceq}) with recourse to the bond-dilution Ising model on the square lattice.
We consider the following Hamiltonian 
\begin{equation}
H = - \sum_{\langle ij \rangle} J_{ij} S_iS_j,
\end{equation}
where $S_i$ stands for the Ising spin taking $\pm 1$, and $J_{ij}$ stands for the random coupling following the distribution functions
\begin{equation}
P_p(J_{i}) = p \delta(J_{ij}-J) + (1-p)\delta(J_{ij}).
\end{equation}
We also define $P_s$, $P_t$ and $P_u$ for each bond.

In random spin system, we need to take the configurational average of $J_{ij}$ to evaluate the free energy.
We often employ the replica method to perform the configurational average.
Instead of the averaged logarithm of the partition function (free energy), we analyze the averaged power following the well-known identity as
\begin{equation}
\left[ \log Z \right] = \lim_{n \to 0} \frac{\left[Z^n\right]-1}{n},
\end{equation}
where $[\cdots]$ expresses the configurational average.
Initially we deal with the replicated system by setting $n$ as a natural number.
At the final step of analysis, we take the limit of $n \to 0$.
We then regard the averaged power of the partition function $[Z^n]$ as the effective partition function written as $Z_n$ (the replicated partition function).

Let us perform the duality analysis with real-space renormalization by dealing with the effective partition function.
The effective partition function consists of the following edge Boltzmann factor as
\begin{equation}
x_{\{S^{\alpha}_{i}\}} = \left[ \prod_{\alpha=1}^n\exp(K \tau_{ij}S^{\alpha}_{i}S^{\alpha}_j) \right],
\end{equation}
where $K=\beta J$, and $\tau_{ij}$ takes $0$ or $1$ and expresses the existence of the interaction.
The superscript $\alpha$ runs from $1$ to $n$ standing for the index of the replicas.
On the other hand the dual edge Boltzmann factor is defined as
\begin{eqnarray}\nonumber
&&x^*_{\{S^{\alpha}_{i}\}} \\ \nonumber
&&= \left(\frac{1}{\sqrt{2}}\right)^n\left[ \prod_{\alpha=1}^n\left({\rm e}^{K \tau_{ij}} + S^{\alpha}_{i}S_j^{\alpha}{\rm e}^{K \tau_{ij}} \right) \right].\\
\end{eqnarray}

Let us take the cluster with four bonds as in Fig. \ref{fig5} to evaluate Eq. (\ref{MCP_cluster}) for $b=1$.
In order to evaluate the renormalized-edge Boltzmann factors, we fix the edge spins to $S_i=1$ on the cluster and sum over the internal spin similarly to the star-triangle transformation.
The renormalized-principal Boltzmann factor is written as 
\begin{equation}
x_0^{(1)} = \left[ \left\{ \sum_{S_0} \prod_{i}\exp(K \tau_{i}S_0) \right\}^n\right],
\end{equation}
where the product runs over $i=p,r,s$ and $t$.
The dual renormalized-principal Boltzmann factor is given by 
\begin{equation}
x_0^{*(1)} = \left[ \left\{ \left(\frac{1 }{4}\right)\sum_{S_0} \prod_{i}\left({\rm e}^{K \tau_{i}}+{\rm e}^{-K \tau_{i}}S_0 \right) \right\}^n\right].
\end{equation}
Taking $n \to 0$ in Eq. (\ref{MCP_cluster}), we obtain the following formula
\begin{eqnarray}\nonumber
&&\left[ \log \left( \frac{ \prod_{i}2\cosh K \tau_{i}}{2\cosh \sum_i K \tau_i}  \right) \left(1 + \prod_i\tanh K \tau_i\right)\right] = 2 \log 2.\\
\end{eqnarray}
In order to identify the location of the bond-percolation thresholds, we consider $K \to \infty$.
We obtain
\begin{equation}
\left[ \log \left(2^{4-\sum_{i}\tau_i-\prod_{i}(1-\tau_{i}) }\left( 1 + \prod_{i} \tau_i \right) \right)\right] = 2 \log 2.\label{Con0}
\end{equation}

First, let us take the homogenous case $p=r=s=t$
Equation (\ref{Con0}) becomes
\begin{equation}
-p^4 - 4p^3(1-p) + 4p(1-p)^3 + (1-p)^4 = 0.
\end{equation}
This implies
\begin{equation}
p^4 + 4p^3(1-p) = 4p(1-p)^3 + (1-p)^4.
\end{equation}

Let us obtain the generic formula for the inhomogenous case (\ref{Ceq}).
From Eq. (\ref{Con0}) we find
\begin{widetext}
\begin{eqnarray}\nonumber
&& prst +\left\{prs(1-t) + pr (1-s) t + p(1-r) st + (1-p) rst \right\}\\ \nonumber
&& \quad - \left\{p(1-r)(1-s)(1-t) + (1-p)r (1-s) (1-t) + (1-p)(1-r) s(1-t)  + (1-p)(1-r)(1-s)t \right\} \\
&& \qquad - (1-p)(1-r)(1-s)(1-t) = 0.\label{PPbefore}
\end{eqnarray}
By simplifying the above equality, we reproduce Eq. (\ref{Ceq}).

We can find the general formula ($\ref{PP2}$) as follows.
The first term in Eq. (\ref{PPbefore}) $prst$ corresponds to $P(ABCD)$ in Eq. (\ref{PP2}).
The following four terms in the first line is $P(ABC|D)$, $P(ACD|B)$, $P(ABD|C)$, and $P(BCD|A)$, respectively.
The remaining terms become $-P(A|B|C|D)$. 
Therefore the above equality (\ref{PPbefore}) suggests the generic formula (\ref{PP2}).

On the other hand, the simple duality as $p^* = 1-p$ reduces Eq. (\ref{PPbefore}) to
\begin{eqnarray}\nonumber
&& p^*r^*s^*t^*+\left\{p^*r^*s^*(1-t^*) + p^*r^* (1-s^*) t^* + p^*(1-r^*) s^*t^* + (1-p^*) r^*s^*t^* \right\}\\ \nonumber
&& \quad - \left\{p^*(1-r^*)(1-s^*)(1-t^*) + (1-p^*)r^* (1-s^*) (1-t^*) + (1-p^*)(1-r^*) s^*(1-t^*)  + (1-p^*)(1-r^*)(1-s^*)t^* \right\} \\
&& \qquad - (1-p^*)(1-r^*)(1-s^*)(1-t^*) = 0.\label{PPbeforedual}
\end{eqnarray}
\end{widetext}
Then the collection of all the terms in the first line becomes $P(\bar{A}\bar{B}\bar{C}\bar{D})$.
Each term in the second line corresponds to $-P(\bar{A}\bar{B}|\bar{C}|\bar{D})$, $-P(\bar{B}\bar{C}|\bar{D}|\bar{A})$, $-P(\bar{D}\bar{A}|\bar{B}|\bar{C})$, and $-P(\bar{C}\bar{D}|\bar{A}|\bar{B})$, respectively.
The last term is nothing but $-P(\bar{A}|\bar{B}|\bar{C}|\bar{D})$.
We thus obtain another generic formula (\ref{PP3}).

\bibliography{paper}

\begin{thebibliography}{22}
\expandafter\ifx\csname natexlab\endcsname\relax\def\natexlab#1{#1}\fi
\expandafter\ifx\csname bibnamefont\endcsname\relax
  \def\bibnamefont#1{#1}\fi
\expandafter\ifx\csname bibfnamefont\endcsname\relax
  \def\bibfnamefont#1{#1}\fi
\expandafter\ifx\csname citenamefont\endcsname\relax
  \def\citenamefont#1{#1}\fi
\expandafter\ifx\csname url\endcsname\relax
  \def\url#1{\texttt{#1}}\fi
\expandafter\ifx\csname urlprefix\endcsname\relax\def\urlprefix{URL }\fi
\providecommand{\bibinfo}[2]{#2}
\providecommand{\eprint}[2][]{\url{#2}}

\bibitem[{\citenamefont{Stauffer and Aharony}(1994)}]{Stauffer1994}
\bibinfo{author}{\bibfnamefont{D.}~\bibnamefont{Stauffer}} \bibnamefont{and}
  \bibinfo{author}{\bibfnamefont{A.}~\bibnamefont{Aharony}},
  \emph{\bibinfo{title}{Introduction To Percolation Theory}}
  (\bibinfo{publisher}{CRC Press}, \bibinfo{year}{1994}),
  \bibinfo{edition}{2nd} ed., ISBN \bibinfo{isbn}{0748402535},
  \urlprefix\url{http://www.worldcat.org/isbn/0748402535}.

\bibitem[{\citenamefont{Kramers and Wannier}(1941)}]{Kramers1941}
\bibinfo{author}{\bibfnamefont{H.~A.} \bibnamefont{Kramers}} \bibnamefont{and}
  \bibinfo{author}{\bibfnamefont{G.~H.} \bibnamefont{Wannier}},
  \bibinfo{journal}{Phys. Rev.} \textbf{\bibinfo{volume}{60}},
  \bibinfo{pages}{252} (\bibinfo{year}{1941}),
  \urlprefix\url{http://link.aps.org/doi/10.1103/PhysRev.60.252}.

\bibitem[{\citenamefont{Wu and Wang}(1976)}]{Wu1976}
\bibinfo{author}{\bibfnamefont{F.~Y.} \bibnamefont{Wu}} \bibnamefont{and}
  \bibinfo{author}{\bibfnamefont{Y.~K.} \bibnamefont{Wang}},
  \bibinfo{journal}{Journal of Mathematical Physics}
  \textbf{\bibinfo{volume}{17}}, \bibinfo{pages}{439} (\bibinfo{year}{1976}),
  \urlprefix\url{http://link.aip.org/link/?JMP/17/439/1}.

\bibitem[{\citenamefont{Wu}(1982)}]{Wu1982}
\bibinfo{author}{\bibfnamefont{F.~Y.} \bibnamefont{Wu}}, \bibinfo{journal}{Rev.
  Mod. Phys.} \textbf{\bibinfo{volume}{54}}, \bibinfo{pages}{235}
  (\bibinfo{year}{1982}),
  \urlprefix\url{http://link.aps.org/doi/10.1103/RevModPhys.54.235}.

\bibitem[{\citenamefont{Nishimori and Ortiz}(2011)}]{Nishimori2011}
\bibinfo{author}{\bibfnamefont{H.}~\bibnamefont{Nishimori}} \bibnamefont{and}
  \bibinfo{author}{\bibfnamefont{G.}~\bibnamefont{Ortiz}},
  \emph{\bibinfo{title}{Elements of Phase Transitions and Critical Phenomena
  (Oxford Graduate Texts)}} (\bibinfo{publisher}{Oxford University Press, USA},
  \bibinfo{year}{2011}), ISBN \bibinfo{isbn}{0199577226}.

\bibitem[{\citenamefont{Wu}(1979)}]{Wu1979}
\bibinfo{author}{\bibfnamefont{F.~Y.} \bibnamefont{Wu}},
  \bibinfo{journal}{Journal of Physics C: Solid State Physics}
  \textbf{\bibinfo{volume}{12}}, \bibinfo{pages}{L645} (\bibinfo{year}{1979}),
  \urlprefix\url{http://stacks.iop.org/0022-3719/12/i=17/a=002}.

\bibitem[{\citenamefont{Ziff et~al.}(2012)\citenamefont{Ziff, Scullard,
  Wierman, and Sedlock}}]{Ziff2012}
\bibinfo{author}{\bibfnamefont{R.~M.} \bibnamefont{Ziff}},
  \bibinfo{author}{\bibfnamefont{C.~R.} \bibnamefont{Scullard}},
  \bibinfo{author}{\bibfnamefont{J.~C.} \bibnamefont{Wierman}},
  \bibnamefont{and} \bibinfo{author}{\bibfnamefont{M.~R.~A.}
  \bibnamefont{Sedlock}}, \bibinfo{journal}{Journal of Physics A: Mathematical
  and Theoretical} \textbf{\bibinfo{volume}{45}}, \bibinfo{pages}{494005}
  (\bibinfo{year}{2012}),
  \urlprefix\url{http://stacks.iop.org/1751-8121/45/i=49/a=494005}.

\bibitem[{\citenamefont{Kesten}(1982)}]{Kesten1982}
\bibinfo{author}{\bibfnamefont{H.}~\bibnamefont{Kesten}},
  \emph{\bibinfo{title}{Percolation theory for mathematicians}}, Progress in
  probability and statistics (\bibinfo{publisher}{Birkh{\"a}user},
  \bibinfo{year}{1982}), ISBN \bibinfo{isbn}{9783764331078}.

\bibitem[{\citenamefont{Scullard and Ziff}(2008)}]{Scullard2008}
\bibinfo{author}{\bibfnamefont{C.~R.} \bibnamefont{Scullard}} \bibnamefont{and}
  \bibinfo{author}{\bibfnamefont{R.~M.} \bibnamefont{Ziff}},
  \bibinfo{journal}{Phys. Rev. Lett.} \textbf{\bibinfo{volume}{100}},
  \bibinfo{pages}{185701} (\bibinfo{year}{2008}),
  \urlprefix\url{http://link.aps.org/doi/10.1103/PhysRevLett.100.185701}.

\bibitem[{\citenamefont{Nakanishi et~al.}(1981)\citenamefont{Nakanishi,
  Reynolds, and Redner}}]{Nakanishi1981}
\bibinfo{author}{\bibfnamefont{H.}~\bibnamefont{Nakanishi}},
  \bibinfo{author}{\bibfnamefont{P.~J.} \bibnamefont{Reynolds}},
  \bibnamefont{and} \bibinfo{author}{\bibfnamefont{S.}~\bibnamefont{Redner}},
  \bibinfo{journal}{Journal of Physics A: Mathematical and General}
  \textbf{\bibinfo{volume}{14}}, \bibinfo{pages}{855} (\bibinfo{year}{1981}),
  \urlprefix\url{http://stacks.iop.org/0305-4470/14/i=4/a=015}.

\bibitem[{\citenamefont{Ohzeki et~al.}(2008)\citenamefont{Ohzeki, Nishimori,
  and Berker}}]{Ohzeki2008}
\bibinfo{author}{\bibfnamefont{M.}~\bibnamefont{Ohzeki}},
  \bibinfo{author}{\bibfnamefont{H.}~\bibnamefont{Nishimori}},
  \bibnamefont{and} \bibinfo{author}{\bibfnamefont{A.~N.}
  \bibnamefont{Berker}}, \bibinfo{journal}{Phys. Rev. E}
  \textbf{\bibinfo{volume}{77}}, \bibinfo{pages}{061116}
  (\bibinfo{year}{2008}),
  \urlprefix\url{http://link.aps.org/doi/10.1103/PhysRevE.77.061116}.

\bibitem[{\citenamefont{Ohzeki}(2009)}]{Ohzeki2009a}
\bibinfo{author}{\bibfnamefont{M.}~\bibnamefont{Ohzeki}},
  \bibinfo{journal}{Phys. Rev. E} \textbf{\bibinfo{volume}{79}},
  \bibinfo{pages}{021129} (\bibinfo{year}{2009}),
  \urlprefix\url{http://link.aps.org/doi/10.1103/PhysRevE.79.021129}.

\bibitem[{\citenamefont{Reynolds et~al.}(1977)\citenamefont{Reynolds, Stanley,
  and Klein}}]{Raynolds1977}
\bibinfo{author}{\bibfnamefont{P.~J.} \bibnamefont{Reynolds}},
  \bibinfo{author}{\bibfnamefont{H.~E.} \bibnamefont{Stanley}},
  \bibnamefont{and} \bibinfo{author}{\bibfnamefont{W.}~\bibnamefont{Klein}},
  \bibinfo{journal}{Journal of Physics C: Solid State Physics}
  \textbf{\bibinfo{volume}{10}}, \bibinfo{pages}{L167} (\bibinfo{year}{1977}),
  \urlprefix\url{http://stacks.iop.org/0022-3719/10/i=8/a=002}.

\bibitem[{\citenamefont{Reynolds et~al.}(1980)\citenamefont{Reynolds, Stanley,
  and Klein}}]{Raynolds1980}
\bibinfo{author}{\bibfnamefont{P.~J.} \bibnamefont{Reynolds}},
  \bibinfo{author}{\bibfnamefont{H.~E.} \bibnamefont{Stanley}},
  \bibnamefont{and} \bibinfo{author}{\bibfnamefont{W.}~\bibnamefont{Klein}},
  \bibinfo{journal}{Phys. Rev. B} \textbf{\bibinfo{volume}{21}},
  \bibinfo{pages}{1223} (\bibinfo{year}{1980}),
  \urlprefix\url{http://link.aps.org/doi/10.1103/PhysRevB.21.1223}.

\bibitem[{\citenamefont{Ziff}(2006)}]{Ziff2006}
\bibinfo{author}{\bibfnamefont{R.~M.} \bibnamefont{Ziff}},
  \bibinfo{journal}{Phys. Rev. E} \textbf{\bibinfo{volume}{73}},
  \bibinfo{pages}{016134} (\bibinfo{year}{2006}),
  \urlprefix\url{http://link.aps.org/doi/10.1103/PhysRevE.73.016134}.

\bibitem[{\citenamefont{Wierman}(1984)}]{Wierman1984}
\bibinfo{author}{\bibfnamefont{J.~C.} \bibnamefont{Wierman}},
  \bibinfo{journal}{Journal of Physics A: Mathematical and General}
  \textbf{\bibinfo{volume}{17}}, \bibinfo{pages}{1525} (\bibinfo{year}{1984}),
  \urlprefix\url{http://stacks.iop.org/0305-4470/17/i=7/a=020}.

\bibitem[{\citenamefont{Feng et~al.}(2008)\citenamefont{Feng, Deng, and
  Bl\"ote}}]{Feng2008}
\bibinfo{author}{\bibfnamefont{X.}~\bibnamefont{Feng}},
  \bibinfo{author}{\bibfnamefont{Y.}~\bibnamefont{Deng}}, \bibnamefont{and}
  \bibinfo{author}{\bibfnamefont{H.~W.~J.} \bibnamefont{Bl\"ote}},
  \bibinfo{journal}{Phys. Rev. E} \textbf{\bibinfo{volume}{78}},
  \bibinfo{pages}{031136} (\bibinfo{year}{2008}),
  \urlprefix\url{http://link.aps.org/doi/10.1103/PhysRevE.78.031136}.

\bibitem[{\citenamefont{Scullard}(2012)}]{Scullard2012a}
\bibinfo{author}{\bibfnamefont{C.~R.} \bibnamefont{Scullard}},
  \bibinfo{journal}{Phys. Rev. E} \textbf{\bibinfo{volume}{86}},
  \bibinfo{pages}{041131} (\bibinfo{year}{2012}),
  \urlprefix\url{http://link.aps.org/doi/10.1103/PhysRevE.86.041131}.

\bibitem[{\citenamefont{Jacobsen and Scullard}(2012)}]{Jacobsen2012}
\bibinfo{author}{\bibfnamefont{J.~L.} \bibnamefont{Jacobsen}} \bibnamefont{and}
  \bibinfo{author}{\bibfnamefont{C.~R.} \bibnamefont{Scullard}},
  \bibinfo{journal}{Journal of Physics A: Mathematical and Theoretical}
  \textbf{\bibinfo{volume}{45}}, \bibinfo{pages}{494003}
  (\bibinfo{year}{2012}),
  \urlprefix\url{http://stacks.iop.org/1751-8121/45/i=49/a=494003}.

\bibitem[{\citenamefont{Scullard and Jacobsen}(2012)}]{Scullard2012b}
\bibinfo{author}{\bibfnamefont{C.~R.} \bibnamefont{Scullard}} \bibnamefont{and}
  \bibinfo{author}{\bibfnamefont{J.~L.} \bibnamefont{Jacobsen}},
  \bibinfo{journal}{Journal of Physics A: Mathematical and Theoretical}
  \textbf{\bibinfo{volume}{45}}, \bibinfo{pages}{494004}
  (\bibinfo{year}{2012}),
  \urlprefix\url{http://stacks.iop.org/1751-8121/45/i=49/a=494004}.

\bibitem[{\citenamefont{{Ohzeki} and {Fujii}}(2012)}]{Ohzeki2012e}
\bibinfo{author}{\bibfnamefont{M.}~\bibnamefont{{Ohzeki}}} \bibnamefont{and}
  \bibinfo{author}{\bibfnamefont{K.}~\bibnamefont{{Fujii}}},
  \bibinfo{journal}{ArXiv e-prints}  (\bibinfo{year}{2012}),
  \eprint{1209.3500}.

\bibitem[{\citenamefont{Onsager}(1944)}]{Onsager1944}
\bibinfo{author}{\bibfnamefont{L.}~\bibnamefont{Onsager}},
  \bibinfo{journal}{Phys. Rev.} \textbf{\bibinfo{volume}{65}},
  \bibinfo{pages}{117} (\bibinfo{year}{1944}),
  \urlprefix\url{http://link.aps.org/doi/10.1103/PhysRev.65.117}.

\end{thebibliography}
\end{document}